\documentclass[twocolumn,prc,superscriptaddress,unsortedaddress,preprintnumbers,amsmath,amssymb]{revtex4}
\usepackage{graphicx}
\usepackage{dcolumn}
\usepackage{bm}
\usepackage{CJK}
\usepackage{CJKulem}
\usepackage{txfonts}
\usepackage[dvipdfmx,bookmarks=true,colorlinks,%
citecolor=blue,linkcolor=blue,anchorcolor=blue,filecolor=blue,urlcolor=blue,%
           ]{hyperref}

\usepackage{mathptmx, courier, pifont}
\usepackage[scaled=0.92]{helvet}
\usepackage[T1]{fontenc}
\usepackage{textcomp}
\usepackage{color}
\usepackage{colordvi}

\def\beq{\begin{equation}}
\def\eeq{\end{equation}}
\def\bea{\begin{eqnarray}}
\def\eea{\end{eqnarray}}

\def\fun#1#2{\lower3.6pt\vbox{\baselineskip0pt\lineskip.9pt
  \ialign{$\mathsurround=0pt#1\hfil##\hfil$\crcr#2\crcr\sim\crcr}}}

\begin{document}
\begin{CJK*} {GBK} {song}
\preprint{}
\title{Breakdown of the tensor component in the Skyrme energy density
    functional}

\author{J. M. Dong}\email[ ]{dongjm07@impcas.ac.cn}\affiliation{Institute of Modern Physics, Chinese
Academy of Sciences, Lanzhou 730000, China} \affiliation{School of
Physics, University of Chinese Academy of Sciences, Beijing 100049,
China}
\author{X. L. Shang}\affiliation{Institute of Modern Physics, Chinese
Academy of Sciences, Lanzhou 730000, China}\affiliation{School of
Physics, University of Chinese Academy of Sciences, Beijing 100049,
China}

\date{\today}

\begin{abstract}
The tensor force, as an important component of strong nuclear force,
generates a variety of intriguing effects ranging from few-body
systems to neutron stars. It is responsible for the nucleon-nucleon
correlation beyond mean-field approximation, and is accordingly
proved to play no role in the standard Skyrme energy density
functionals in the present work. Therefore, the Skyrme's original
tensor interaction that is extensively-employed presently is
invalid. As an alternative strategy, we introduced a central
interaction, i.e., the $\bm{\sigma }_{1}\cdot \bm{\sigma }_{2}$
term, to improve the description of experimental single-particle
structure, and to address its effect, we established two Skyrme
interactions IMP1 and IMP2 complemented by the calibrated
charge-violating interactions. The central $\bm{\sigma }_{1}\cdot
\bm{\sigma }_{2}$ interaction turns out to substantially improve the
description of shell evolution in Sn isotopes and $N=82$ isotones.
\end{abstract}
\maketitle

\section{Introduction}\label{intro}\noindent
Exotic nuclei far away from the $\beta$-stability line exhibit many
novel and striking features, which are at the exciting forefront in
the contemporary nuclear physics and open an intriguing test ground
for nuclear models~\cite{Nature1,Nature2,Nature3}. A fundamental
understanding of how the shell structure evolves from stable to
exotic nuclei, and how it impacts the relevant nuclear structure
properties and the $r$-process of nucleosynthesis, is one of the
primary challenges of modern experimental and theoretical nuclear
physics. The tensor force, as a non-central component of the
realistic nuclear force, is generally believed to play a significant
role, in particular in the shell evolution~\cite{Otsuka1} and
spin-isospin excitation such as the Gamow-Teller and spin-dipole
states~\cite{BCL}. Moreover, the short-range correlation (SRC)
(creating a high-momentum tail), dominated by the strong tensor
force between unlike nucleons, has far-reaching impact on areas as
diverse as the nuclear structure~\cite{Science2008,Science2014},
transport properties and superfluidity of dense nuclear
matter~\cite{Dong-SRC1,BAL}, neutron star cooling~\cite{Dong-SRC1}
and the EMC effect~\cite{Nature2018,EMC2}, highlighting the
fundamental importance of the tensor force.

A zero-range tensor potential was present in the original Skyrme
interaction~\cite{SKY1}, and its role in the evolution of the
nuclear single-particle levels was discussed firstly by Stancu {\it
et al} ~\cite{Stancu}. Over the past decade, the interest for the
tensor force was revived~\cite{SO1,SO2,SO3,SO4}, driven by the
production of a large amount of new exotic nuclei following the
development of modern radioactive beam facilities and experimental
detectors~\cite{TF-Review}. In this work, combined with the
framework of an $ab$ initio method, we demonstrate that such a
widely-used tensor force is invalid, and then we propose a new
scheme to replace this counterfeit tensor force so as to improve the
description of single-particle levels.

\section{Why is the tensor component of the Skyrme interaction invalid?}\label{intro}\noindent
Our starting point is the tensor force in microscopic many-body
approaches. For infinite homogeneous nuclear matter system, the
momentum distribution around the Fermi level significantly departs
from the typical profile of a degenerate ideal Fermi gas as the
result of the SRC~\cite{SRC11,SRC12,SRC13,SRC14}, where the SRC is
predominantly caused by the short-range repulsion core and the
tensor interaction~\cite{Science2014} (Some authors distinguish the
tensor correlation from the short-range correlation, but here we do
not). The tensor force acting only on spin-triplet states of a two
nucleon system, provides a strong attraction in $T=0$ channel, i.e.,
the $^{3}SD_1$ channel, which is responsible for the binding of
deuteron and is indispensible for the binding of symmetric matter at
saturation density. Here $^{2S+1}L_J$ denotes the state of relative
motion between two nucleons, where $L$, $S$ and $J$ are the relative
orbital angular momentum, the spin and total angular momentum,
respectively. As the AV18 interaction is able to reproduce the
deuteron properties, its tensor-force component is well-defined and
is consistent with the deuteron structure. The momentum
distributions (occupation probability versus momentum $k$) achieved
from $ab$ initio Brueckner theory with the bare AV18 interaction, as
present in Fig. 3 of Ref.~\cite{AV18}, indicate the symmetric matter
system deviates from the right-angle distribution more sizeably than
the pure neutron matter system, which means the correlation in the
former is much stronger than that in the latter~\cite{Science2008}.
This is because the most dominant tensor interaction stems from the
off-diagonal $^{3}SD_1$ neutron-proton coupling that is powerful for
symmetric matter but vanishes for pure neutron matter, which
probably indicates that the neutron-rich nuclei can be better
described by mean-field or energy-density functional approaches.

We attempt to derive an effective tensor potential supplemented to
zero-range Skyrme interactions. It is well-known that, the
simplification of the Fourier transform
\begin{equation}
V_{k^{\prime }k} = \int e^{-i\bm{k}' \cdot \bm{r}}V(\bm{r}) e^{i\bm{k}\cdot \bm{r}%
}d\bm{r}, \label{VKK}
\end{equation}
via low-momentum expansion to the 2nd-order of $\bm {k}$ or $\bm
{k}'$ gives the zero-range momentum-dependent Skyrme effective
interaction, where $\bm{r}=\bm{r}_1-\bm{r}_2$ is the distance
between two nucleons. The operator $\bm{k}=(\overrightarrow{\nabla
}_{1}-\overrightarrow{\nabla }_{2})/(2i)$ acts on the right and
$\bm{k}^{\prime }=-(\overleftarrow{\nabla
}_{1}-\overleftarrow{\nabla } _{2})/(2i)$ acts on the left. We
employ the well-defined tensor force of
\begin{equation}
V_{T}(\bm{r})=f_{T}(r)S_{12}(\bm{r}),\text{ }S_{12}(\bm{r})=\frac{3(%
\bm{\sigma }_{1}\cdot \bm{r})(\bm{\sigma }_{2}\cdot \bm{r})}{\bm{r}^{2}}-%
\bm{\sigma }_{1}\cdot \bm{\sigma }_{2},
\end{equation}
in coordinate space, and then the final Skyrme-type interaction in
momentum space requires to be achieved via the Fourier transform of
Eq. (\ref{VKK}). The Skyrme's original tensor force was introduced
in an unreasonable way, because the tensor-force operator $S_{12}$
in momentum space but with a $r$-dependent strength, i.e.,
$f_{T}(r)S_{12}(\bm{k})$, is applied as a starting point.

However, unlike the central force, the integral of Eq. (\ref{VKK})
for the tensor-force component is difficult or even impossible to
work out since the interaction is anisotropic and $e^{i\bm{k}\cdot
\bm{r}}$ is not the eigenstate of $V_{T}(\bm{r})$. To this end, we
separate the spin wavefunction $\chi _{\sigma }$ from the full
wavefunction $\phi (\bm{r},\bm{\sigma },\bm{\tau })$, and then
calculate $V_{T,k^{\prime }k}=\int \chi _{\sigma _{1}}^{\dag }\chi
_{\sigma _{2}}^{\dag }e^{-i\bm{k'}\cdot
\bm{r}}V_{T}(\bm{r})e^{i\bm{k}\cdot \bm{r}}\chi _{\sigma _{2}}\chi
_{\sigma _{1}}d\bm{r}$ which is exactly the interaction matrix
element in the framework of Brueckner theory. By employing the
angular momentum algebra, its explicit expression is written as the
sum of contributions from various partial-wave channels via
\begin{eqnarray}
V_{T,k^{\prime }k} &=&\underset{J,m_{J}}{\sum }\underset{L,m_{L}}{\sum }%
\underset{L^{\prime },m_{L^{\prime }}}{\sum }\underset{m_{s},m_{s^{\prime }}}%
{\sum }\left( 4\pi \right) ^{2}i^{L-L^{\prime }}C_{\frac{1}{2}m_{s_{1}}\frac{%
1}{2}m_{s_{2}}}^{1m_{s}}C_{Lm_{L}1m_{s}}^{Jm_{J}}  \notag \\
&&C_{\frac{1}{2}m_{s_{1}}\frac{1}{2}m_{s_{2}}}^{1m_{s}^{\prime
}}C_{L^{\prime }m_{L}^{\prime }1m_{s}^{\prime
}}^{Jm_{J}}Y_{L^{\prime }}^{m_{L^{\prime }}\ast
}(\widehat{\bm{k}}^{\prime
})Y_{L}^{m_{L}}(\widehat{\bm{k}})\bigg[2\delta _{L^{\prime
},J}\delta _{L,J}+
\notag \\
&&\delta _{L^{\prime },J-1}\delta _{L,J-1}\frac{2-2J}{2J+1}+\delta
_{L^{\prime },J-1}\delta _{L,J+1}\frac{6\sqrt{J(J+1)}}{2J+1}+  \notag \\
&&\delta _{L^{\prime },J+1}\delta
_{L,J-1}\frac{6\sqrt{J(J+1)}}{2J+1}-\delta
_{L^{\prime },J+1}\delta _{L,J+1}\frac{2(J+2)}{2J+1}\bigg]  \notag \\
&&\cdot \int j_{L^{\prime }}(k^{\prime }r)f_{T}(r)j_{L}(kr)r^{2}dr,
\label{A3}
\end{eqnarray}
where $Y_{L}^{m_{L}}(\widehat{\bm{k}})$ is the spherical harmonic
and $j_L(kr)$ is the $L$th-order spherical Bessel function.

\begin{table}[h]
\caption{\label{table1} The individual reasons for the drop of each
tensor channel in Skyrme-type interaction, $ \bigcirc $: exactly
zero according to Eq. (\ref{A3}); $\otimes $: zero due to the
orthogonality between spherical harmonics for off-diagonal matrix
element; $\boxtimes $: the order of $\mathcal{O}(\bm{k}^2)$ beyond
standard Skyrme forces; $\bigtriangleup $: can be reabsorbed into
existing Skyrme components. }
\begin{ruledtabular}
\begin{tabular}{cccc}
Channel & $V_T^{(L,L)}$  & $V_T^{(L,L+2)}$ & $V_T^{(L+2,L+2)}$\\
\hline
$^3SD_1$  & $ \bigcirc $ & $\otimes $ & $\boxtimes $ \\
$^3P_0$  & $\bigtriangleup $ & -- & -- \\
$^3P_1$  & $\bigtriangleup $ & -- & -- \\
$^3D_2$  & $\boxtimes $ & -- & -- \\
$^3PF_2$  & $\bigtriangleup $ & $\otimes $ & $\boxtimes $ \\
\end{tabular}
\end{ruledtabular}
\end{table}

For uniform nuclear matter, the off-diagonal matrix element, i.e.,
$V_{T,k^{\prime }k}(^{3}L_{J}-^{3}(L+2)_{J})$ plays no role for the
binding energy directly in the mean-field approximation due to the
orthogonality between $Y_{L^{\prime }}^{m_{L^{\prime }}\ast }(\widehat{\bm{k}'})$ and $Y_{L}^{m_L}(%
\widehat{\bm{k}})$ for $L \ne L'$, but it is essential for the
correlation between nucleons particularly the $^3SD_1$ tensor
channel. In finite nuclei, such off-diagonal matrix elements are
also expected to have no role under the mean-field approximation
which can be interpreted as the vanishing monopole component that
averages over all orientations of $\bm{k}$. The diagonal element
$V_{T,k'k}(^3S_1-^3S_1)$ is exactly zero in terms of Eq. (\ref{A3}),
whereas $V_{T,k^{\prime }k}\left( ^{3}D_{1}-^{3}D_{1}\right)
\varpropto \int j_{2}(k^{\prime }r)V_{T}(r)j_{2}(kr)dr$ is in the
order of $\bm{k}^4$ that is beyond the standard Skyrme interactions.
The non-zero monopole component of triplet-odd tensor terms,
$V_{T,k'k}(^3P_J-^3P_J) \varpropto \int j_{1}(k^{\prime
}r)V_{T}(r)j_{1}(kr)dr \sim \bm{k}^{\prime }\cdot \bm{k}$, can be
reabsorbed into the existing $t_2$-$x_2$ term in Skyrme
interactions. The four types of reasons why each tensor channel is
dropped in Skyrme-type interaction are summarized in
Table~\ref{table1}. In short, one cannot introduce the tensor force
in the standard Skyrme-Hartree-Fock calculations additionally, and
the tensor-force-induced correlation between nucleons leading to
Fermi surface depletion of dense matter systems, as shown in Fig. 3
of Ref.~\cite{AV18}, is completely unavailable since it is beyond
the mean-field approximation. As a consequence, the role of the
tensor force in Skyrme energy density functionals is eventually
clarified. As pointed out in Ref.~\cite{single1}, a zero-range
implementation of the tensor interaction in the Skyrme-type
interaction is problematic, and the role of correlations is required
to be understood.

\begin{table}[h]
\caption{\label{table2} IMP1 and IMP2 Skyrme parameter sets.
$a_{\text{CSB}}$ is determined by the Brueckner-Hartree-Fock method
with AV18 interaction~\cite{Dong2017}. The bottom grouping shows the
corresponding properties of symmetric nuclear matter, including the
binding energy per nucleon $B/A$, the incompressibility $K_{\infty
}$ and the isoscalar effective mass $m_{\infty }^*/m$ at saturation
density $\rho_0$, the symmetry energy $J$ and its slope $L$ at
$\rho_0$. The maximum neutron star mass $M_{\text{max}}$, and a
radius $R$ for a canonical neutron star, along with the neutron skin
thickness and symmetry energy coefficient of $^{208}$Pb are also
listed.}
\begin{ruledtabular}
\begin{tabular}{cccc}
parameter & IMP1  & IMP2 \\
\hline $t_0$ (MeV fm$^3$) & -2380.9896  &   -2349.5057 \\
$t_1$ (MeV fm$^5$) & 486.7908  & 506.4560 \\
$t_2$ (MeV fm$^4$) & -351.2604 & -347.7734  \\
$t_3$ (MeV fm$^{3(1+\gamma)}$) & 12614.2135  & 12265.1049 \\
$x_0$  & 1.0866 & 1.1719 \\
$x_1$  & -0.7449 & -0.8106 \\
$x_2$  &-1.0000  &-1.0000  \\
$x_3$  & 1.8933 &  2.0937\\
$W_0$ (MeV fm$^5$) & 127.3999 & 122.8769 \\
$\gamma$  & 1/6 & 1/6 \\
$a_{\text{CSB}}$ (MeV$ $fm$^{-3}$)  & -1.0513  & -1.0513  \\
$a_{\text{exc}}$  & 0.30  & 0.28  \\
$U_s $  & 0  & -2218.5231  \\
$U_t $  & 0  & 1851.2276 \\
$J^2 $ term  & No  & Yes  \\
\hline
$\rho_0$ (fm$^{-3}$) & 0.160  &  0.160 \\
$B/A$ (MeV) & -16.0  &  -16.0 \\
$K_{\infty }$ (MeV) & 236  &  239 \\
$m_{\infty }^*/m$ & 0.65  &  0.64 \\
$J$ (MeV) & 32.1  &  31.8 \\
$L$ (MeV) & 44  &  43 \\
$M_{\text{max}}$ ($M_{\odot }$) & 2.1  &  2.1 \\
$R_{1.4 M_{\odot }}$ (km) & 11.7  &  11.8 \\
$\Delta R_{np}(^{208}\text{Pb})$ (fm) & 0.15  &  0.14 \\
$a_{\text{sym}}(^{208}\text{Pb})$ (MeV) & 23.9  &  23.9 \\

\end{tabular}
\end{ruledtabular}
\end{table}

\section{A new strategy to improve shell evolution within Skyrme functionals}

A question arises immediately: the tensor force, such as the strong
attractive $^3SD_1$ channel responsible for deuteron structure, does
not directly contribute to the standard Skyrme interaction, can we
introduce the other mechanism to reproduce the experimental
single-particle structure? The answer to this question is provided
below.

The $\bm{\sigma }_{1}\cdot \bm{\sigma }_{2}$ term, as a central
force, is expected to appear in bare nuclear interaction such as the
one-pion exchange potential $V_{\text{OPEP}}(\bm{r})=f_{\pi
}^{2}m_{\pi }\bm{\tau }_{1}\cdot \bm{\tau }_{2}\left[
f(r)S_{12}+\bm{\sigma }_{1}\cdot \bm{\sigma }_{2}e^{-m_{\pi
}r}/(3m_{\pi }r)\right] $~\cite{TF-Review}, which is regarded as the
remainder of spin- and isospin-dependent meson-nucleon couplings
after the tensor force is separated. The non-zero $(\bm{\sigma
}_{1}\cdot \bm{\sigma }_{2})\left( 1-P^{r}P^{\sigma }P^{\tau
}\right) $ can be achieved for four spin-isospin-parity ($S$, $T$,
$P$) channels, including (1,0,+), (0,1,+), (0,0,-) and (1,1,-). Here
$P^{r}$, $P^{\sigma }$ and $P^{\tau}$ are the position-, spin- and
isospin-exchange operators. It should be noticed that the (1,0,+)
and (0,0,-) channels give the same kind of contribution to the
density functional ($\varpropto \bm{\sigma }_{1}\cdot \bm{\sigma
}_{2}(1-\delta _{q_{1}q_{2}})$); and also the (0,1,+) and (1,1,-)
channels give the same kind of contribution ($\varpropto \bm{\sigma
}_{1}\cdot \bm{\sigma }_{2}$). Accordingly, we only take into
account the (1,0,+) and (1,1,-) channels with new-defined strengthes
for the sake of discussion, and here the approximation of
$P_{12}^{\tau }=\delta _{q_{1}q_{2}}$ is dropped. The effective
zero-range interaction corresponding to such a central $\bm{\sigma
}_{1}\cdot \bm{\sigma }_{2}$ component is written as
\begin{equation}
V_{\bm{\sigma }}=\bm{\sigma }_{1}\cdot \bm{\sigma }%
_{2}\left\{ \frac{U_{s}}{4}\left[ \bm{k}^{\prime 2}\delta (\bm{r})+\delta (%
\bm{r})\bm{k}^{2}\right] +\frac{U_{t}}{2}\bm{k}^{\prime }\delta (\bm{r})%
\bm{k}\right\} ,  \label{A4}
\end{equation}
with the coupling strength $U_s$ and $U_t$ as free parameters. This
component only modifies the $\mathcal{H}_{\text{sg}}$ term that is
generated by the tensor coupling with spin and gradient, and it is
expressed as
\begin{eqnarray}
\mathcal{H}_{sg} &=&\left[ -\frac{1}{16}\left( t_{1}x_{1}+t_{2}x_{2}\right) -%
\frac{1}{16}\left( U_{s}+U_{t}\right) \right] \bm{J}^{2}+  \notag \\
&&\left[ \frac{1}{16}\left( t_{1}-t_{2}\right)
+\frac{1}{16}U_{s}\right] \left(
\bm{J}_{n}^{2}+\bm{J}_{p}^{2}\right) ,
\end{eqnarray}
where $\bm{J}$ is the spin density. Correspondingly, the strength
for the spin-orbit potential for spherical nuclei is given by
\begin{equation}
U_{\text{s.o.}}^{(q)}=\frac{W_{0}}{2r}\left( 2\frac{d\rho _{q}}{dr}+\frac{%
d\rho _{q^{\prime }}}{dr}\right) +\left( \alpha \frac{\bm{J}_{q}}{r}%
+\beta \frac{\bm{J}_{q^{\prime }}}{r}\right),
\end{equation}
where $q (q')$ denotes the like (unlike) particles. The first term
on the right-hand side arises from the Skyrme spin-orbit
interaction, while the second term includes the contributions of a
central exchange term and the $\bm{\sigma }_{1}\cdot \bm{\sigma
}_{2}$ term, i.e., $\alpha=\alpha_c + \alpha_\sigma$, $\beta=\beta_c
+ \beta_\sigma$, which are expressed as
\begin{eqnarray}
\alpha _{c} &=&\frac{1}{8}\left( t_{1}-t_{2}\right)
-\frac{1}{8}\left(
t_{1}x_{1}+t_{2}x_{2}\right) ,  \\
\beta _{c} &=&-\frac{1}{8}\left( t_{1}x_{1}+t_{2}x_{2}\right) ,  \\
\alpha _{\sigma } &=&-\frac{1}{8}U_{t}, \\
\beta _{\sigma } &=&-\frac{1}{8}\left( U_{s}+U_{t}\right) .
\end{eqnarray}
Therefore, the onset of the central $\bm{\sigma }_{1}\cdot
\bm{\sigma }_{2}$ interaction is responsible for the spin-orbit
splitting.

\begin{figure}[htbp]
\begin{center}
\includegraphics[width=0.45\textwidth]{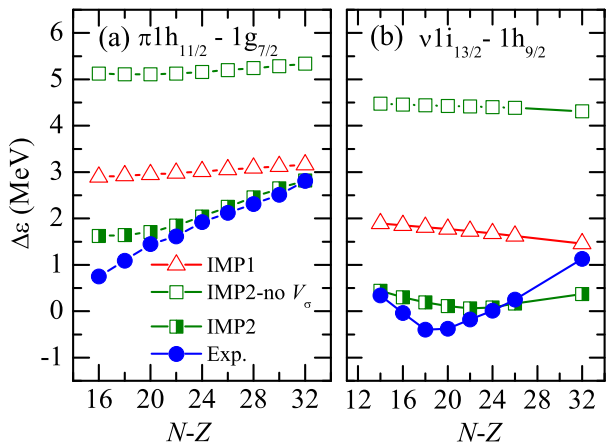}
\caption{Energy differences between the $1h_{11/2}$ and $1g_{7/2}$
single-proton states along Sn isotopes, between $1h_{9/2}$ and
$1i_{13/2}$ single-neutron states along the $N=82$ isotones
predicted by the IMP2 (with and without $\bm{\sigma }_{1}\cdot
\bm{\sigma }_{2}$ interaction) and IMP1 forces, compared with
experimental data~\cite{XXXX}.}\label{fig1}
\end{center}
\end{figure}

To assess the effect of the $\bm{\sigma }_{1}\cdot \bm{\sigma }_{2}$
interaction on the single-particle structure, we establish new
Skyrme interactions. The empirical properties of infinite nuclear
matter around the saturation density, and the well-determined
ground-state observables of a selected set of closed-shell nuclei,
are taken advantage of to accurately calibrate Skyrme
parametrizations through a chi-square minimization procedure. The
equation of state of pure neutron matter given by the realistic
calculation AV18+$\delta v$+UIX$^*$ of Akmal {\it et al.}~\cite{PNM}
is included in the fits for its good description of neutron star
properties. The spin-orbit strength is adjusted by choosing the
single-particle energy splittings of neutron $3p$ in $^{208}$Pb, and
neutrons and protons $2p$ in $^{56}$Ni. Furthermore, the symmetry
energy coefficient $a_{\text{sym}}$ of $^{208}$Pb that has been
achieved with the wealth of measured data on nuclear masses and
decay energies~\cite{Dong2018}, is employed as an important
isovector benchmark for a reliable construction of density
functionals for the first time. The center-of-mass corrections to
the binding energy of finite nuclei is obtained by
$E_{\text{c.m.}}=0.75(45A^{-1/3}-25A^{-2/3})$ MeV as introduced in
Ref.~\cite{Ecm}.

The effective two-body charge-symmetry breaking (CSB) nuclear
interaction from the Brueckner-Hartree-Fock approach~\cite{Dong2017}
for {\it asymmetric} matter, in completely contrast to
Ref.~\cite{Colo2018}, and the renormalized coulomb coupling
coefficient $e_{0}^2=e^2 (1+a_{\text{exc}} Z^{-2/3})$
phenomenologically embodying many complicated
corrections~\cite{Dong2019}, are introduced aiming to reproduce the
experimental Coulomb displacement energy (CDE) of mirror nuclei (the
binding-energy difference between two members of a mirror pair) and
the excitation energy of isobaric analog state (IAS). Accordingly,
it reconciles our knowledge of the symmetry energy around the
saturation density with the IAS energy in a heavy nucleus. The
insignificant charge-independent-breaking interaction is dropped in
the present study.

A parameter set that neglects the $\bm{J}^2$ term just as the
widely-used SLy4~\cite{SLY4} is built, referred to as the `Institute
of Modern Physics 1' (IMP1) interaction. Then we provide the other
parameter set `IMP2' that includes the $\bm{J}^2$ term together with
$\bm{\sigma }_{1}\cdot \bm{\sigma }_{2}$ interaction of
Eq.~(\ref{A4}), keeping the merits of IMP1 for global nuclear
properties, and the single-particle energy differences between
proton $1g_{7/2}$ and $1h_{11/2}$ for $^{120}$Sn and $^{132}$Sn, and
between neutron $1h_{9/2}$ and $1i_{13/2}$ for $^{146}$Gd and
$^{132}$Sn serve as additional calibrations. Both two parameter
sets, and the resulting properties of bulk nuclear matter, neutron
star and finite nuclei, are listed in Table~\ref{table2}. The rms
deviation of the binding energy and charge rms radius are 1.2 MeV
(1.3 MeV) and 0.035 fm (0.037 fm) for IMP1 (IMP2) force
respectively, which are satisfactory (at the $1\%$ level or better).
Intriguingly, the new interactions are quite successful in the
description of a variety of CDE of mirror nuclei that results from
the charge-violating interactions. For example, for the mirror pair
of $^{48}$Ni-$^{48}$Ca, the theoretical results of 67.00 MeV (67.30
MeV) within the IMP1 (IMP2) force, is in excellent agreement with
AME2016 value of 67.28(48) MeV~\cite{DATA2016}. In addition, the
maximum neutron star mass, and a radius for a $1.4 M_{\odot}$
canonical neutron star, are compatible with astrophysical
observations. Therefore, these two parametrizations enable us to
investigate the global nuclear properties reliably.

We mainly concentrate upon the effect of the central $\bm{\sigma
}_{1}\cdot \bm{\sigma }_{2}$ interaction. In Fig.~\ref{fig1}, the
isospin-dependence of the energy differences $\Delta \varepsilon $
between $1h_{11/2}$ and $1g_{7/2}$ single-proton states outside the
$Z=50$ core along the Sn isotopes, and between $1h_{9/2}$ and
$1i_{13/2}$ single-neutron states outside the $N=82$ core along the
$N=82$ isotones, are displayed as a function of the neutron excess
$N-Z$, calculated with the IMP1 and IMP2 forces in comparison with
the experimental measurements. The interaction IMP1, and also the
IMP2 without the $\bm{\sigma }_{1}\cdot \bm{\sigma }_{2}$ component,
fail to reproduce the experimental trend qualitatively, whereas the
full IMP2 force improves the agreement with experimental data
considerably without destroying existing description of nuclei,
being attributed to the additional $\bm{\sigma }_{1}\cdot \bm{\sigma
}_{2}$ term of Eq. (\ref{A4}). In other words, the $\bm{\sigma
}_{1}\cdot \bm{\sigma }_{2}$ interaction has a robust and systematic
effect on the single-particle levels of nuclei and hence shell
evolutions.

\section{Summary and outlook}
Based on the derived tensor force in momentum space via partial-wave
expansion, we demonstrated that the tensor force plays no role in
Skyrme density functionals, and hence the invalidity of the
widely-employed tensor force of the Skyrme interactions. The tensor
force we discussed here is well-defined and is responsible for the
deuteron structure and nucleon-nucleon correlation. Then we proposed
a new strategy to improve the shell evolution, i.e., by introducing
the central $\bm{\sigma }_{1}\cdot \bm{\sigma }_{2}$ interaction. To
examine our strategy, two Skyrme interactions (IMP1 and IMP2) were
built, where both the effective CSB interaction and the renormalized
coulomb coupling coefficient are introduced. The shell evolution in
Sn isotopes and $N =82$ isotones can be well reproduced.

Essentially, the physical picture that the nucleons move as
independent particles subject to a mean field generated by all the
other nucleons, is apparently oversimplified. Consequently, the
self-consistent mean-field calculations cannot be expected to
accurately describe experimental single-particle energies and hence
the isospin-dependence of shell structure. The underlying mechanisms
beyond the mean-field approximation that account for the
single-particle structure, such as the tensor force, the short-range
repulsion core and the particle vibration coupling, are still
subjects of great challenge. Their effects on the single-particle
structure are considered to be phenomenologically embodied into the
$\bm{\sigma }_{1}\cdot \bm{\sigma }_{2}$ interaction through
parameters fitting. Therefore, the parameter sets that is applicable
to shell evolutions cannot be anticipated to be valid any more for
excited states such as Gamow-Teller and spin-dipole states. In the
long term, how to include the nucleon-nucleon correlations that are
induced by tensor force and short-range repulsion, based on
energy-density functionals, is an essential scientific problem in
nuclear physics, and work along this line is underway.

\section*{Acknowledgement}This work was supported by the National Natural Science Foundation
of China under Grants No. 11775276, by the Youth Innovation
Promotion Association of Chinese Academy of Sciences.

\end{CJK*}

\end{document}